\documentclass[conference,twocolumn,10pt]{IEEEtran}

\pdfoutput=1
\usepackage[cmex10]{amsmath}
\usepackage{amsfonts,amssymb}
\usepackage{verbatim}
\usepackage{color}
\usepackage{multirow} % For table column to span multiple rows

\usepackage{cite}
\ifCLASSINFOpdf
  \usepackage[pdftex]{graphicx}
  \graphicspath{{graphics/}}
  \DeclareGraphicsExtensions{.pdf,.jpeg,.png}
\else
  \usepackage[dvips]{graphicx}
  \graphicspath{{graphics/}}
  \DeclareGraphicsExtensions{.eps}
\fi

\usepackage{array}
\usepackage{mdwmath}
\usepackage{mdwtab}
\DeclareMathOperator{\erf}{erf}

\begin{document}
\bibliographystyle{IEEEtran}
\title{Channel Impulse Responses in Diffusive Molecular Communication with Spherical Transmitters}

\author{\IEEEauthorblockN{Adam Noel$^{\dagger\ast}$, Dimitrios Makrakis$^{\dagger}$, and Abdelhakim Hafid$^{\ast}$}
\IEEEauthorblockA{$^{\dagger}$School of Electrical Engineering and Computer Science,
	University of Ottawa \\ $^{\ast}$Department of Computer Science and Operations Research,
	University of Montreal}}


% Calculus
\newcommand{\x}{x}
\newcommand{\y}{y}
\newcommand{\z}{z}

% Units
\newcommand{\second}{\textnormal{s}}
\newcommand{\metre}{\textnormal{m}}

% Words
\newcommand{\threeD}{\textnormal{3D}}
\newcommand{\oneD}{\textnormal{1D}}
\newcommand{\pass}{\textnormal{PA}}
\newcommand{\absorb}{\textnormal{AB}}
\newcommand{\mol}{\textnormal{mol}}
\newcommand{\PTA}{\textnormal{PTA}}

\newcommand{\dev}{\Delta_\textnormal{dev}}

% Chemical parameters
\newcommand{\kth}[1]{k_{#1}}

% Physical parameters
\newcommand{\M}{M}
\newcommand{\T}{T}
\newcommand{\At}{A(\tx{})}
\newcommand{\Dx}[1]{D_{#1}}
\newcommand{\Nx}[1]{N_{#1}}
\newcommand{\Nxtavg}[2]{\overline{\Nx{}}_\textnormal{#2}\left(#1\right)}
\newcommand{\NxAvg}[1]{\overline{\Nx{}}_\textnormal{#1}}
\newcommand{\Cxtavg}[2]{\overline{C}_\textnormal{#2}\left(#1\right)}
\newcommand{\Nxt}[2]{{\Nx{}}_\textnormal{#2}\left(#1\right)}

% Functions
\newcommand{\EXP}[1]{\exp\left(#1\right)}
\newcommand{\ERF}[1]{\erf\left(#1\right)}
\newcommand{\ERFC}[1]{\mathrm{erfc}\left(#1\right)}

% Arbitrary constants

% Simulator variables
\newcommand{\Vol}[1]{V_{#1}}
\newcommand{\subV}[1]{S_{#1}}
\newcommand{\dt}[1]{\Delta\tx{\textnormal{#1}}}
\newcommand{\prop}[1]{a_{#1}}
\newcommand{\molNum}[1]{U_{#1}}
\newcommand{\rx}{\textnormal{RX}}
\newcommand{\xI}{x_\textnormal{i}}
\newcommand{\xF}{x_\textnormal{f}}
\newcommand{\rad}[1]{\vec{r}_{#1}}
\newcommand{\rrx}{r_\textnormal{RX}}
\newcommand{\rtx}{r_\textnormal{TX}}
\newcommand{\Vrx}{V_\textnormal{RX}}
\newcommand{\Vtx}{V_\textnormal{TX}}
\newcommand{\dist}{d}
\newcommand{\distInt}{d_\textnormal{int}}
\newcommand{\tx}[1]{t_{#1}}
\newcommand{\tpeak}{t_{\mathrm{peak}}}
\newcommand{\kx}[1]{k_{#1}}
\newcommand{\wx}[1]{w_{#1}}
\newcommand{\thresh}{\xi}

\newcommand{\weight}[1]{w_{#1}}

\newcommand{\smM}{m}

\newtheorem{definition}{Definition}

% Counter to allow double column equations
\newcounter{mytempeqncnt}

\maketitle

\begin{abstract}
Molecular communication is an emerging paradigm for systems that rely on the release of molecules as information carriers. Communication via molecular diffusion is a popular strategy that is ubiquitous in nature and very fast over distances on the order of a micron or less. Existing closed-form analysis of the diffusion channel impulse response generally assumes that the transmitter is a point source. In this paper, channel impulse responses are derived for spherical transmitters with either a passive or absorbing receiver. The derived channel impulse responses are in closed-form for a one-dimensional environment and can be found via numerical integration for a three-dimensional environment. The point transmitter assumption (PTA) is formally defined so that its accuracy can be measured in comparison to the derived spherical transmitter impulse responses. The spherical transmitter model is much more accurate than the PTA when the distance between a transmitter and its receiver is small relative to the size of the transmitter. The derived results are verified via microscopic particle-based simulations using the molecular communication simulation platform AcCoRD (Actor-based Communication via Reaction-Diffusion). A spherical transmitter variation where molecules are released from the surface of a solid sphere is also considered via simulation.
\end{abstract}

\section{Introduction}

Molecular communication (MC) is a physical layer design strategy where molecules are used as information carriers. It is ubiquitous for communication in biological systems, but over the past decade it has also received growing attention for synthetic communication in fluid environments where conventional communication strategies are infeasible or undesirable; see \cite{Farsad2016}. One MC approach that is commonly found in biological cells is free diffusion, where molecules propagate via collisions with each other; see \cite[Ch.~3]{Alberts}. From a conventional wireless communications perspective, communication via diffusion can be perceived as analogous to broadcast communication (as described in \cite{Atakan2010}), since the transmitter has no direct influence on the trajectory of molecules once they are released and molecules might be observed by any receiver capable of detecting them.

While there is a firm theoretical foundation for the modeling of diffusion-based processes, closed-form channel impulse responses are only available with simplifying assumptions and under specific environment geometries; see \cite{Crank1979}. One common assumption that has been widely applied in the MC literature is that a transmitter is modeled as a point source of molecules; see \cite[Ch.~5]{Nakano2013c}. We formally define this here as the \emph{point transmitter assumption} (PTA). In practice, a transmitter (and correspondingly its container of molecules) will occupy physical space, so it is of interest to account for the initial physical distribution of molecules when they begin to diffuse. This was identified as an open problem in \cite{Farsad2016}. We anticipate that modeling a transmitter as a finite volume will substantially improve the accuracy of the channel impulse response when the transmitter is close to its intended receiver.

In this paper, we formulate the channel impulse response at a spherical receiver due to molecules diffusing from a spherical transmitter. The transmitter is a virtual boundary that releases molecules by initializing them uniformly within the sphere. The receiver can be either an absorbing surface or a passive observer. Both one-dimensional (1D) and three-dimensional (3D) environments are considered, since channel impulse responses from a point transmitter are available in closed form for both absorbing and passive receivers in 1D and 3D. Our contributions can be summarized as follows:
\begin{enumerate}
	\item We derive the spherical transmitter channel impulse responses for passive and absorbing receivers in 1D and 3D environments. The impulse responses can be obtained in closed form in 1D, and are obtained via numerical integration in 3D.
	\item We define the point transmitter assumption (PTA), which facilitates the measurement of its deviation from the spherical transmitter model.
	\item We use simulations to verify all spherical transmitter impulse responses and compare them with the PTA. For a passive receiver, we also compare with the uniform concentration assumption (which we coined in \cite{Noel2013b}), which effectively treats the receiver as a point observer. The spherical transmitter impulse responses are shown to initially be much more accurate than the approximations when the transmitter is close to the receiver.
	\item We briefly use simulations to consider the ``solid'' spherical transmitter, where molecules are released by initializing them on its surface. In this scenario, it is still more accurate to use the spherical transmitter impulse response for the virtual boundary transmitter than to use the PTA.
\end{enumerate}

We conduct microscopic particle-based simulations using our AcCoRD simulator (Actor-based Communication via Reaction-Diffusion), which is in active development. Source code and executables are available on Github; see \cite{Noel2016}.

Existing papers that have considered volume transmitters include \cite{Chou2015b,Yilmaz2014a} and related works by the same authors. \cite{Chou2015b} uses a mesoscopic model where the environment is divided into subregions called \emph{voxels}, such that the transmitter can occupy one or more voxels, and channel responses are derived in the Laplace domain. \cite{Yilmaz2014a} simulates spherical transmitters but does not account for finite transmitter volume in the corresponding analytical model.

The rest of this paper is organized as follows. Section~\ref{sec_model} describes the system model and presents the channel impulse responses for point sources. Section~\ref{sec_volume_tx} derives the spherical transmitter channel impulse responses and defines the point transmitter assumption. We verify our analysis with numerical results and simulations in Section~\ref{sec_results}, and conclude in Section~\ref{sec_concl}.

\section{System Model and Analytical Preliminaries}
\label{sec_model}

We consider a transmitter with radius $\rtx$ releasing $\Nx{}$ molecules omni-directionally into an unbounded 1D or 3D environment with uniform temperature and viscosity. Thus, the transmitter is a segment of length $2\rtx$ in 1D and a sphere in 3D. The released molecules are initially distributed uniformly throughout the transmitter. Local molecule concentrations are low enough to assume that the molecules diffuse with constant diffusion coefficient $\Dx{}$. These molecules are observed by a receiver that is centered at a distance $\dist$ from the center of the transmitter and has radius $\rrx$. If the receiver is active, then we consider a perfectly-absorbing surface that removes and counts molecules as they arrive, i.e., the absorption rate is $\kx{} \to \infty$. If the receiver is passive, then it has no impact on molecule behavior but is able to count the number of molecules within its virtual boundary at any instant. Finally, we assume that diffusion is the only phenomenon affecting molecule behavior in the propagation environment (except for absorption at the absorbing receiver's surface). We leave the integration of other realistic molecular phenomena, such as bulk fluid flow or the potential for other chemical reactions, for future work.

We define the channel impulse response $\Nxtavg{t}{RX}$ as the number of molecules \emph{expected} at the receiver at time $\tx{}$, given that $\Nx{}$ molecules are instantaneously released by the transmitter at time $\tx{}=0$ (i.e., we assume that the period of molecule release is negligibly small). In the following subsections, we present known channel impulse responses for a point transmitter. These impulse responses are needed to derive the channel impulse responses of the corresponding spherical transmitters in Section~\ref{sec_volume_tx}. We also briefly comment on the expected signal peak times, which we will use in Section~\ref{sec_results} to normalize time when assessing the accuracy of the point transmitter assumption.

\subsection{1D Channel Impulse Responses Due to a Point Source}

If the environment is 1D, then the channel impulse response of the \emph{absorbing} receiver due to a point source is given by \cite[Eq.~(7)]{Farsad2016}
\begin{equation}
\Nxtavg{t}{RX}|^\absorb_\oneD = \Nx{}\ERFC{\frac{\dist-\rrx}{\sqrt{4\Dx{}\tx{}}}},
\label{absorbing_response_1D}
\end{equation}
where we use the ``$\absorb$'' superscript for ``absorbing''. 
$\ERFC{x} = 1 - \ERF{x}$ is the complementary error function (from \cite[Eq.~(8.250.4)]{Gradshteyn2007}), and the error function is \cite[Eq.~(8.250.1)]{Gradshteyn2007}
\begin{equation}
\label{APR12_32}
\ERF{x} = \frac{2}{\sqrt{\pi}}\int\limits_0^x \EXP{-y^2} \mathrm{d}y.
\end{equation}

The expected \emph{point} concentration for the 1D \emph{passive} receiver is \cite[Eq.~(3.6)]{Crank1979}
\begin{equation}
\Cxtavg{t}{point}|^\pass_\oneD = \frac{\Nx{}}{\sqrt{4\pi \Dx{}
		\tx{}}}\EXP{-\frac{\dist^2}{4\Dx{}\tx{}}},
\label{passive_response_point_1D}
\end{equation}
where we use the ``$\pass$'' superscript to denote ``passive''.

We can integrate (\ref{passive_response_point_1D}) over the receiver line segment in a manner analogous to our derivation of the impulse response at a 3D \emph{Cartesian box} receiver in \cite[Eq.~(22)]{Noel2013b} to arrive at
\begin{equation}
\Nxtavg{t}{RX}|^\pass_\oneD = \frac{\Nx{}}{2}\left(\ERF{\frac{\rrx+\dist}{2\sqrt{\Dx{}\tx{}}}}
- \ERF{\frac{\dist-\rrx}{2\sqrt{\Dx{}\tx{}}}}\right).
\label{passive_response_exact_1D}
\end{equation}

It is common to assume that the molecule concentration throughout a passive receiver at any time is uniform, which is justified if the receiver is sufficiently far from the transmitter, i.e., if $\dist \gg \rrx$ (as we demonstrated in \cite{Noel2013b}). If we do so, then we can scale (\ref{passive_response_point_1D}) by the receiver volume $|\Vrx| = 2\rrx$ and write
\begin{equation}
\Nxtavg{t}{RX}|^\pass_\oneD = \frac{\rrx\Nx{}}{\sqrt{\pi \Dx{}
		\tx{}}}\EXP{-\frac{\dist^2}{4\Dx{}\tx{}}}.
\label{passive_response_UCA_1D}
\end{equation}

\subsection{3D Channel Impulse Responses Due to a Point Source}

If the environment is 3D, then the channel impulse response of the \emph{absorbing} receiver due to a point source is given by \cite[Eq.~(23)]{Yilmaz2014b}
\begin{equation}
\Nxtavg{t}{RX}|^\absorb_\threeD = \frac{\Nx{}\rrx}{\dist}\ERFC{\frac{\dist-\rrx}{\sqrt{4\Dx{}\tx{}}}}.
\label{absorbing_response}
\end{equation}

Eq.~(\ref{absorbing_response})
describes the \emph{total} number of molecules that have been absorbed by time $\tx{}$. For the \emph{passive} receiver, the expected \emph{point} concentration $\Cxtavg{t}{point}$ due to a point source is \cite[Eq.~(4.28)]{Crank1979}
\begin{equation}
\Cxtavg{t}{point}|^\pass_\threeD = \frac{\Nx{}}{(4\pi \Dx{}
	\tx{})^{3/2}}\EXP{-\frac{\dist^2}{4\Dx{}\tx{}}}.
\label{passive_response_point}
\end{equation}

We previously integrated (\ref{passive_response_point}) over a sphere to derive the impulse response of the passive spherical receiver as \cite[Eq.~(27)]{Noel2013b}
\begin{align}
\Nxtavg{t}{RX}|^\pass_\threeD = &\; \frac{\Nx{}}{2}\left[\ERF{\frac{\rrx-
		\dist}{2\sqrt{\Dx{}\tx{}}}} +
\ERF{\frac{\rrx+\dist}{2\sqrt{\Dx{}\tx{}}}}\right] \nonumber
\\
& +
\frac{\Nx{}}{\dist}\sqrt{\frac{\Dx{}\tx{}}{\pi}}
\bigg[\EXP{-\frac{(\dist+\rrx)^2}{4\Dx{}\tx{}}} \nonumber \\
& - \EXP{-\frac{(\dist-\rrx)^2}{4\Dx{}\tx{}}}\bigg].
\label{passive_response_exact}
\end{align}

If we assume that the passive receiver is sufficiently far from the transmitter, then the simplified 1D channel impulse response comes directly from (\ref{passive_response_point}) as
\begin{equation}
\Nxtavg{t}{RX}|^\pass_\threeD = \frac{\Nx{}|\Vrx|}{(4\pi \Dx{}
	\tx{})^{3/2}}\EXP{-\frac{\dist^2}{4\Dx{}\tx{}}}.
\label{passive_response_UCA}
\end{equation}

\subsection{Signal Peak Time}

One of the tests in Section~\ref{sec_results} will be to measure the relative deviation of the point transmitter assumption over time for a range of distances between the transmitter and receiver. However, the time delay until the expected arrival of molecules at the receiver is a function of the distance $\dist$, so we will find it convenient to normalize time by the expected signal peak time $\tpeak$ at each distance. For a passive receiver, the expected channel impulse response peak times in 1D and 3D due to a point transmitter can be shown to be
\begin{equation}
\tpeak|^\pass_\oneD = \frac{\dist^2}{2\Dx{}},\qquad \tpeak|^\pass_\threeD = \frac{\dist^2}{6\Dx{}},
\label{tpeak_passive}
\end{equation}
respectively, when using the uniform concentration assumption. For an absorbing receiver, the channel impulse responses (\ref{absorbing_response_1D}) and (\ref{absorbing_response}) are non-decreasing. However, it is common to measure the \emph{rate} of molecule absorption over time, and this is expected to be greatest at time
\begin{equation}
\tpeak|^\absorb = \frac{(\dist-\rrx)^2}{6\Dx{}},
\label{tpeak_absorbing}
\end{equation}
which can be shown to apply in both 1D and 3D when the transmitter is a point.

\section{Channel Impulse Responses for Spherical Transmitters}
\label{sec_volume_tx}

\begin{figure*}[!t]
	% ensure that we have normalsize text
	\normalsize
	% Store the current equation number.
	\setcounter{mytempeqncnt}{\value{equation}}
	% Set the equation number to one less than the one
	% desired for the first equation here.
	% The value here will have to changed if equations
	% are added or removed prior to the place these
	% equations are referenced in the main text.
	\setcounter{equation}{14}
	\begin{align}
	\Nxtavg{t}{RX}|_\oneD^{\Vtx,\pass} = &\, \frac{\Nx{}}{2\rtx}\Bigg\{\sqrt{\frac{\Dx{}\tx{}}{\pi}}
	\Bigg[\EXP{-\frac{(\xF+\rrx)^2}{4\Dx{}\tx{}}} - \EXP{-\frac{(\xF-\rrx)^2}{4\Dx{}\tx{}}}
	- \EXP{-\frac{(\xI+\rrx)^2}{4\Dx{}\tx{}}} \nonumber \\
	& + \EXP{-\frac{(\xI-\rrx)^2}{4\Dx{}\tx{}}}\Bigg] + \frac{1}{2}\Bigg[(\xF + \rrx)\ERF{\frac{\xF + \rrx}{2\sqrt{\Dx{}\tx{}}}} \nonumber \\
	& - (\xI + \rrx)\ERF{\frac{\xI + \rrx}{2\sqrt{\Dx{}\tx{}}}}
	- (\xF - \rrx)\ERF{\frac{\xF - \rrx}{2\sqrt{\Dx{}\tx{}}}} 
	+ (\xI - \rrx)\ERF{\frac{\xI - \rrx}{2\sqrt{\Dx{}\tx{}}}}\Bigg]\Bigg\}
	\label{volumeCIR_passive_1D}
	\end{align}
	\begin{align}
	\Nxtavg{t}{RX}|_\oneD^{\Vtx,\absorb} = &\, \frac{\Nx{}}{2\rtx}\Bigg[(\xF-\xI) + 2\sqrt{\frac{\Dx{}\tx{}}{\pi}}
	\left(\EXP{-\frac{(\xI-\rrx)^2}{4\Dx{}\tx{}}} - \EXP{-\frac{(\xF-\rrx)^2}{4\Dx{}\tx{}}}\right) \nonumber \\
	& - (\xF-\rrx)\ERF{\frac{\xF-\rrx}{2\sqrt{\Dx{}\tx{}}}} + (\xI-\rrx)\ERF{\frac{\xI-\rrx}{2\sqrt{\Dx{}\tx{}}}}\Bigg]
	\label{volumeCIR_absorbing_1D}
	\end{align}
	% Restore the current equation number.
	\setcounter{equation}{\value{mytempeqncnt}}
	% IEEE uses as a separator
	\hrulefill
	% The spacer can be tweaked to stop underfull vboxes.
	\vspace*{4pt}
\end{figure*}

In this section, we derive the channel impulse responses when the transmitter is a virtual spherical source. Generally, the channel impulse response due to such a source is defined as follows. The release of $\Nx{}$ molecules means that $\Nx{}$ molecules are uniformly distributed over the transmitter volume $\Vtx$, i.e., the density of molecules throughout the transmitter is $\Nx{}/|\Vtx|$, where $\Vtx$ could be defined in 1D or 3D. Due to the spherical symmetry of the receiver, the expected observation of a molecule that is initialized at \emph{any} point outside of the receiver is only a function of the distance from that point to the receiver's surface. Thus, the expected channel impulse response of a volume source is found by the corresponding volume integration of the point source channel impulse response.
Given the point source channel impulse response $\Nxtavg{t}{RX}$, the volume source channel impulse response $\Nxtavg{t}{RX}|^{\Vtx}$ is
\begin{equation}
\Nxtavg{t}{RX}|^{\Vtx} = \frac{1}{|\Vtx|} \int\limits_{\Vtx} \Nxtavg{t}{RX} \mathrm{d}\Vtx,
\label{volumeCIR_general}
\end{equation}
which can be solved numerically for any of the point source channel impulse responses presented in Section~\ref{sec_model}. We emphasize that (\ref{volumeCIR_general}) would \emph{not apply} if the receiver were not spherically symmetric, such that the expected observation of a molecule would depend on its precise initial location and not only on its initial distance from the receiver.

In the following subsections, we discuss solving (\ref{volumeCIR_general}) in 1D and 3D environments. For the 1D environment, the channel impulse responses are derived in closed form. For the 3D environment, closed-form solutions are not readily available and we briefly comment on one of the challenges. Finally, we formally define the point transmitter assumption, where we assume that a volume transmitter can be approximated as a point.

\subsection{Line Transmitter in 1D}
\label{sec_1D_tx}

In the 1D environment, the transmitter is a line segment. Thus, the expected channel impulse response is found by solving
\begin{equation}
\Nxtavg{t}{RX}|^{\Vtx}_\oneD = \frac{1}{2\rtx} \int\limits_{-\rtx}^{\rtx} \Nxtavg{t}{RX}|_{\oneD}\, \mathrm{d}c,
\label{volumeCIR_1D}
\end{equation}
where $c$ is the distance from the center of the transmitter to an arbitrary point in $\Vtx$.  If we define the vectors $\rad{c}$ and $\rad{\dist}$ relative to the center of the transmitter, such that $|\rad{c}| = c$ and $|\rad{\dist}| = \dist$ (i.e., $\rad{\dist}$ is the vector from the center of the transmitter to the center of the receiver), then the distance from the center of the \emph{receiver} to an arbitrary point in $\Vtx$, $\distInt$, is
\begin{equation}
\distInt = |\rad{c} - \rad{\dist}| = d+c,
\label{integration_distance_1D}
\end{equation}
and to solve (\ref{volumeCIR_1D}) we replace $\dist$ with $\distInt$ in the appropriate equation for $\Nxtavg{t}{RX}|_{\oneD}$, i.e., in (\ref{absorbing_response_1D}) or (\ref{passive_response_exact_1D}) for the absorbing or passive receiver, respectively.

\addtocounter{equation}{2} % Account for the double column equations here.
We can solve (\ref{volumeCIR_1D}) in closed form for both types of receivers. It can be shown that the expected channel impulse responses at a 1D passive receiver and at a 1D absorbing receiver due to a line transmitter are in (\ref{volumeCIR_passive_1D}) and (\ref{volumeCIR_absorbing_1D}) at the top of the following page, respectively, where
\begin{equation}
\xI = \dist - \rtx, \qquad \xF = \dist + \rtx,
\end{equation}
and we use the integral \cite[Eq.~(5.41)]{Gradshteyn2007}
\begin{equation}
\int\ERF{ax}\mathrm{d}x = x\ERF{ax} + \frac{1}{a\sqrt{\pi}}\EXP{-a^2x^2}.
\label{integration_erf}
\end{equation}

Due to their verbosity, it is challenging to gain intuition from (\ref{volumeCIR_passive_1D}) and (\ref{volumeCIR_absorbing_1D}). However, both impulse responses are \emph{exact} and they show the importance of the precise location of the transmitter (via $\xI$ and $\xF$).

\subsection{Spherical Transmitter in 3D}
\label{sec_3D_tx}

In the 3D environment, the transmitter is a sphere. Thus, the expected channel impulse response is found by solving
\begin{equation}
\Nxtavg{t}{RX}|^{\Vtx}_\threeD = \frac{1}{|\Vtx|} \int\limits_{0}^{\rtx}\int\limits_{0}^{2\pi}\int\limits_{0}^{\pi} \Nxtavg{t}{RX}|_{\threeD}c^2\sin\theta \mathrm{d}\theta \mathrm{d}\phi\mathrm{d}c,
\label{volumeCIR_3D}
\end{equation}
where $c$, as in the 1D case, is the distance from the center of the transmitter to an arbitrary point in $\Vtx$.  The distance from the center of the \emph{receiver} to an arbitrary point in $\Vtx$ is
\begin{equation}
\distInt = |\rad{c} - \rad{\dist}| = \sqrt{c^2 + \dist^2 - 2c\dist\cos\phi\sin\theta},
\label{integration_distance}
\end{equation}
and to solve (\ref{volumeCIR_3D}) we must replace $\dist$ with $\distInt$ in the appropriate equation for $\Nxtavg{t}{RX}|_{\threeD}$, i.e., in (\ref{absorbing_response}) or (\ref{passive_response_exact}) for the absorbing or passive receiver, respectively.

The main challenge to solving (\ref{volumeCIR_3D}) in closed form is that both (\ref{absorbing_response}) and (\ref{passive_response_exact}) have $\dist$ terms that are not squared. This means that the trigonometric terms in $\distInt$ remain inside a square root, and approximations are needed to integrate (\ref{volumeCIR_3D}) with respect to $\theta$ and $\phi$. Accurate approximations have not yet been identified so they are outside the scope of this work. In this paper, we only consider solving (\ref{volumeCIR_3D}) numerically in MATLAB.
%\begin{equation}
%\int\limits_0^{\rtx} \frac{1}{c}\EXP{-Ac^2 - Bc}\mathrm{d}c,
%\end{equation}

\subsection{Point Transmitter Assumption}

In \cite{Noel2013b}, we defined the common \emph{uniform concentration assumption} (UCA) for a receiver, where it is assumed that the molecule concentration throughout a passive receiver at any instant is uniform. Here, we make a similar definition for a point transmitter:

\begin{definition}[Point Transmitter Assumption]
	The \emph{point transmitter assumption} (PTA) assumes that all molecules released by a transmitter are initialized at a single point.
\end{definition}

Even though the PTA has also been a common assumption throughout the molecular communication literature, we establish that the it replaces the integration over $\Vtx$ in (\ref{volumeCIR_general}) with its magnitude $|\Vtx|$. Specifically, the PTA simplifies the volume transmitter impulse response in (\ref{volumeCIR_general}) as
\begin{equation}
\Nxtavg{t}{RX}|^{\Vtx} \approx \Nxtavg{t}{RX}\big|_\PTA^{\Vtx} = \frac{|\Vtx|}{|\Vtx|}\Nxtavg{t}{RX} = \Nxtavg{t}{RX}.
\end{equation}

The definition of the PTA enables us to measure its accuracy. In Section~\ref{sec_results}, we will measure the relative deviation of the PTA. For a given channel impulse response approximation $\Nxtavg{t}{RX}\big|_\textnormal{PTA}^{\Vtx}$, its percent relative deviation $\dev$ is
\begin{equation}
\dev = \frac{\Nxtavg{t}{RX}\big|_\textnormal{PTA}^{\Vtx} - \Nxtavg{t}{RX}|^{\Vtx}}{\Nxtavg{t}{RX}|^{\Vtx}} \times 100\%.
\end{equation}

\section{Simulation and Numerical Results}
\label{sec_results}

In this section, we verify the channel impulse responses derived for spherical transmitters in Section~\ref{sec_volume_tx} via simulations and numerical evaluation. We consider both passive and absorbing receivers in two environments (one that is 1D and one that is 3D). Unless otherwise noted, the two environments have the same parameters as summarized in Table~\ref{table_model_param}. We consider a transmitter and a receiver with radii $\rtx=\rrx=1\,\mu\metre$, which is on the order of the size of a bacterial cell. The diffusion coefficient of released molecules is $\Dx{}=10^{-9}\,\metre^2/\second$. The transmitter and receiver are either centered $2\,\mu\metre$ apart (i.e., they are adjacent) or $5\,\mu\metre$ apart. The transmitter releases $10^3$ molecules at time $\tx{}=0$. The microscopic simulation time step $\dt{sim}$ is $0.1\,\metre\second$ when the receiver is passive and $1\,\mu\second$ when the receiver is absorbing. The smaller time step for simulations with the absorbing receiver is needed to accurately simulate the absorption behavior.

\begin{table}[!tb]
	\centering
	\caption{Simulation system parameters for both 1D and 3D environments (unless otherwise noted).}
	
	{\renewcommand{\arraystretch}{1.4}
		\begin{tabular}{l|c|c||c}
			\hline
			\bfseries Parameter & \bfseries Symbol & \bfseries Units& \bfseries Value \\ \hline \hline
			Transmitter Radius & $\rrx$ &	$\mu\metre$
			& 1 \\ \hline
			Receiver Radius & $\rrx$ &	$\mu\metre$
			& 1  \\ \hline
			Molecules Released & $\Nx{}$ & $\mol$
			& $10^3$ \\ \hline
			Distance to Receiver & $\dist{}$ & $\mu\metre$
			& $\{2,5\}$ \\ \hline
			Diffusion Coefficient
			& $\Dx{}$ & $\metre^2/\second$
			& $10^{-9}$ \\ \hline
			Passive Time Step & $\dt{sim}$ & $\metre\second$
			& 0.1 \\ \hline
			Absorbing Time Step & $\dt{sim}$ & $\mu\second$
			& $1$ \\ \hline
			\# of Realizations & - & -
			& $\ge100$ \\ \hline
		\end{tabular}
	}
	\label{table_model_param}
\end{table}

All particle-based simulations are averaged over at least $100$ independent realizations using release v0.5 of the AcCoRD simulator \cite{Noel2016}, where the 3D environment is truly unbounded and the 1D environment is a long rectangular pipe with dimensions $1\,\metre\metre\times1\,\mu\metre\times1\,\mu\metre$ and having the receiver in the middle. In each time step $\dt{sim}$, all released molecules diffuse. If the receiver is absorbing, then the trajectory of each molecule is checked in each time step for crossing the surface of the receiver.

\begin{figure}[!tb]
	\centering
	\includegraphics[width=\linewidth]{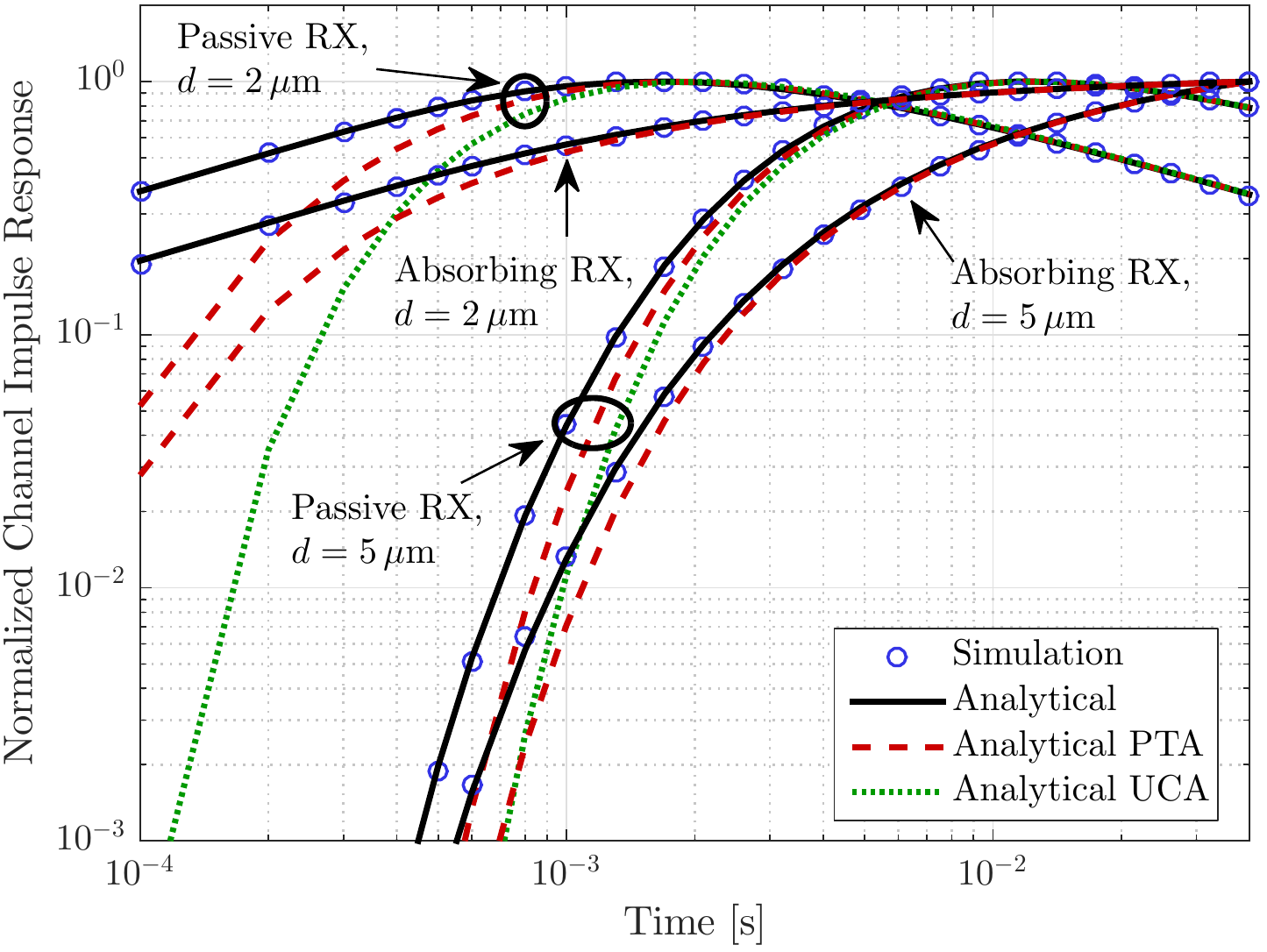}
	\caption{Passive and absorbing channel impulse responses in the 1D environment as a function of time. All curves are normalized to the maximum value of the corresponding analytical curve (solid black line), as listed in Table~\ref{table_norm}.}
	\label{fig_1D_CIR}
\end{figure}

In Fig.~\ref{fig_1D_CIR}, we plot the channel impulse responses of the receivers in the 1D environment. The simulations of the passive receiver are compared with the derived spherical transmitter channel impulse response in (\ref{volumeCIR_passive_1D}), the point transmitter approximation (PTA) in (\ref{passive_response_exact_1D}), and the combination of the PTA and the uniform concentration assumption (UCA) found using (\ref{passive_response_UCA_1D}). The simulations of the absorbing receiver are compared with the derived spherical impulse response in (\ref{volumeCIR_absorbing_1D}) and the PTA in (\ref{absorbing_response_1D}). All curves are normalized by the peak values of the corresponding analytical curves (i.e., solid black lines), as listed in Table~\ref{table_norm}, and are shown on a log-log scale to easily observe deviations in the curves at early times and when a low number of molecules are observed. The only exception to the parameters listed in Table~\ref{table_model_param} is that the simulation time step for the absorbing receiver centered $2\,\mu\metre$ from the transmitter is further reduced to $\dt{sim} = 0.1\,\mu\second$ for accuracy.

\begin{table}[!tb]
	\centering
	\caption{Maximum values used to normalize curves in Figs.~\ref{fig_1D_CIR} and \ref{fig_3D_CIR}.}
	
	{\renewcommand{\arraystretch}{1.4}
		\begin{tabular}{l|c|c|c||c|c}
			\hline
			\bfseries Receiver & \bfseries Symbol & \bfseries Units& \bfseries $\dist$ & \bfseries 1D & \bfseries 3D \\ \hline \hline
			Passive & $\NxAvg{RX}|^{\Vtx,\pass}$ & $\mol$ & $2\,\mu\metre$
			& 242.3 & 38.85 \\ \hline
			Passive & $\NxAvg{RX}|^{\Vtx,\pass}$ & $\mol$ & $5\,\mu\metre$
			& 96.69 & 24.67 \\ \hline
			Absorbing & $\NxAvg{RX}|^{\Vtx,\absorb}$ & $\mol$ & $2\,\mu\metre$
			& 911 & 456 \\ \hline
			Absorbing & $\NxAvg{RX}|^{\Vtx,\absorb}$ & $\mol$ & $5\,\mu\metre$
			& 655 & 1309 \\ \hline
		\end{tabular}
	}
	\label{table_norm}
\end{table}

Fig.~\ref{fig_1D_CIR} shows that the derived spherical transmitter impulse responses are very accurate for both the passive and absorbing receivers over the entire range of time simulated. The approximations initially underestimate the number of molecules at the receiver but become more accurate with increasing time. Applying the UCA with the PTA for the passive receiver, i.e., using a point-to-point model, is noticeably less accurate than only applying the PTA. As expected, the accuracy of the approximations is much poorer when $\dist = 2\,\mu\metre$, i.e., when the transmitter is adjacent to the receiver. Nevertheless, even the point-to-point model is accurate for the passive receiver at this distance when the impulse response is at its peak. We claim that the approximations are sufficient for long-term behavior but they do not accurately model the receiver signal as molecules first arrive from the transmitter.

\begin{figure}[!tb]
	\centering
	\includegraphics[width=\linewidth]{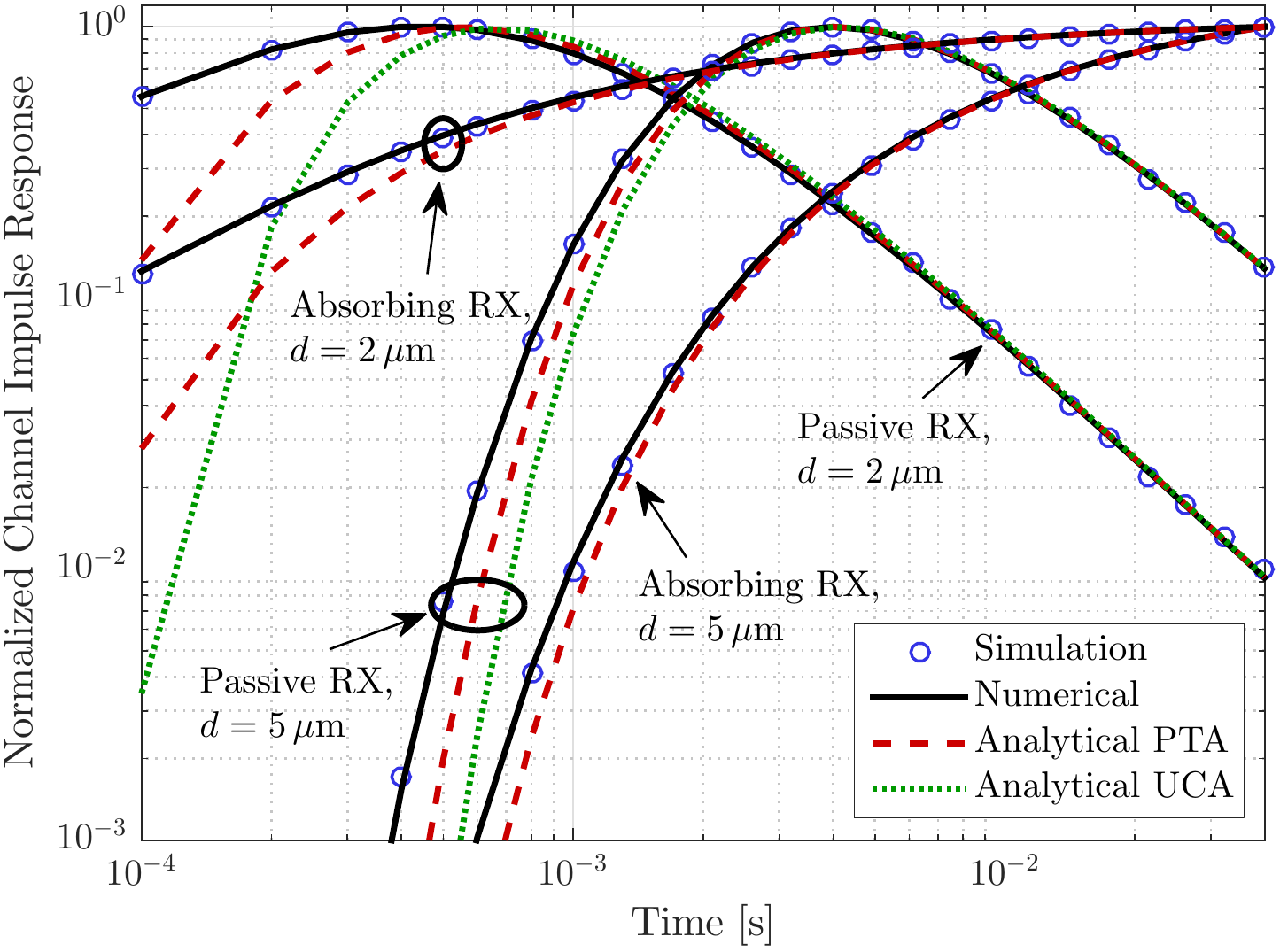}
	\caption{Passive and absorbing channel impulse responses in the 3D environment as a function of time. All curves are normalized to the maximum value of the corresponding analytical curve (solid black line), as listed in Table~\ref{table_norm}.}
	\label{fig_3D_CIR}
\end{figure}

Fig.~\ref{fig_3D_CIR} is analogous to Fig.~\ref{fig_1D_CIR} and plots the channel impulse responses of the receivers in the 3D environment. The spherical transmitter impulse responses are determined by numerically solving the integration in (\ref{volumeCIR_3D}) for all receivers. The PTA for the passive receiver is found via (\ref{passive_response_exact}), and the PTA is combined with the UCA using (\ref{passive_response_UCA}). The PTA for the absorbing receiver is found via (\ref{absorbing_response}). The only exception to the parameters listed in Table~\ref{table_model_param} is that the number of molecules released by the transmitter when $\dist = 5\,\mu\metre$ is increased to $10^4$ so that more molecules can be observed.

The results in Fig.~\ref{fig_3D_CIR} are consistent with our observations from Fig.~\ref{fig_1D_CIR}. Even though we did not derive the spherical transmitter channel impulse responses in closed form, the numerical integration of (\ref{volumeCIR_3D}) agrees with the simulations and is more accurate than the approximations for low values of $\tx{}$. The accuracy of the approximations is worse when the transmitter is closer to the receiver.

\begin{figure}[!tb]
	\centering
	\includegraphics[width=\linewidth]{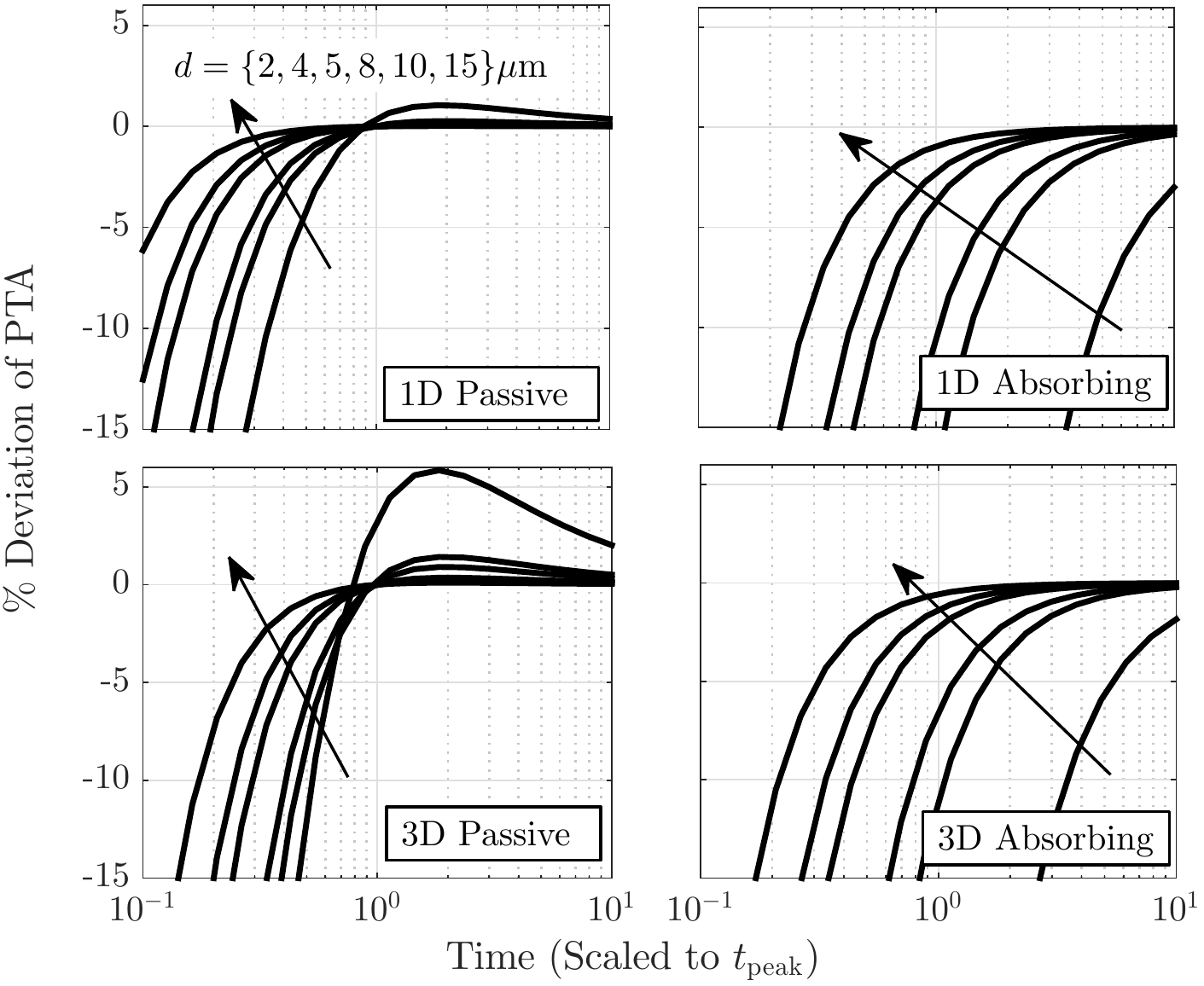}
	\caption{$\%$ deviation of the PTA $\dev$ as a function of (dimensionless) time, for increasing distance $\dist = \{2,4,5,8,10,15\}\mu\metre$ following the arrows as shown. Time is scaled relative to the expected peak time $\tpeak$ of the corresponding channel impulse response using the PTA. All subplots have the same axis scales.}
	\label{fig_pta}
\end{figure}

To get a better sense of the sensitivity of the PTA as a function of distance, we measure its percent deviation $\dev$ from the spherical transmitter impulse response for different distances as a function of time in Fig.~\ref{fig_pta}. We consider both the absorbing and passive receivers in 1D and 3D environments, and consider the distances $\dist =\{2,4,5,8,10,15\}\mu\metre$. Time is normalized for each curve by its corresponding $\tpeak$ in (\ref{tpeak_passive}) or (\ref{tpeak_absorbing}). We see that the deviations of the approximation generally improve with increasing distance, and in fact the exact shape of the transmitter matters less. For a passive receiver, the PTA initially underestimates the impulse response but overestimates it after $\tpeak$. This overestimation is as much as $6\%$ for the 3D receiver when $d=2\,\mu\metre$. Otherwise, all deviations at the passive receiver are less than $5\%$ for all time $\tx{} > 0.7\tpeak$. For an absorbing receiver, the PTA always underestimates the signal, and it improves more slowly than for the passive receiver; for $d<5\,\mu\metre$ the deviation is still greater than $10\%$ when $\tx{}=\tpeak$.

Finally, we briefly consider a variation of the spherical transmitter model, where the transmitter is a solid impenetrable sphere and the molecules that it releases are initialized uniformly over its surface. The analysis to correctly determine the corresponding channel impulse response is outside the scope of this work, but we are interested in assessing the accuracy of the virtual transmitter model in this scenario. We simulated this scenario with a 3D absorbing receiver using the same system parameters as for Fig.~\ref{fig_3D_CIR}, and compared the results with the numerical integration of (\ref{volumeCIR_3D}) and the PTA in (\ref{absorbing_response}). At both distances, (\ref{volumeCIR_3D}) and (\ref{absorbing_response}) initially underestimate the simulation. This makes intuitive sense, since the molecules initialized on the half of the transmitter closest to the receiver will arrive sooner than when predicted by our model, where the molecules are distributed within $\Vtx$. Analogously, (\ref{volumeCIR_3D}) and (\ref{absorbing_response}) eventually overestimate the simulation, since the molecules initialized on the half of the transmitter further from the receiver will take longer to reach the receiver than when predicted. The accuracy of our virtual transmitter model is much better when $\dist = 5\,\mu\metre$ than when $\dist = 2\,\mu\metre$, and at both distances solving (\ref{volumeCIR_3D}) is still more accurate than using the PTA in (\ref{absorbing_response}). While the solid transmitter variation merits further study, the virtual transmitter model may be a sufficient approximation for the channel impulse response.

\begin{figure}[!tb]
	\centering
	\includegraphics[width=\linewidth]{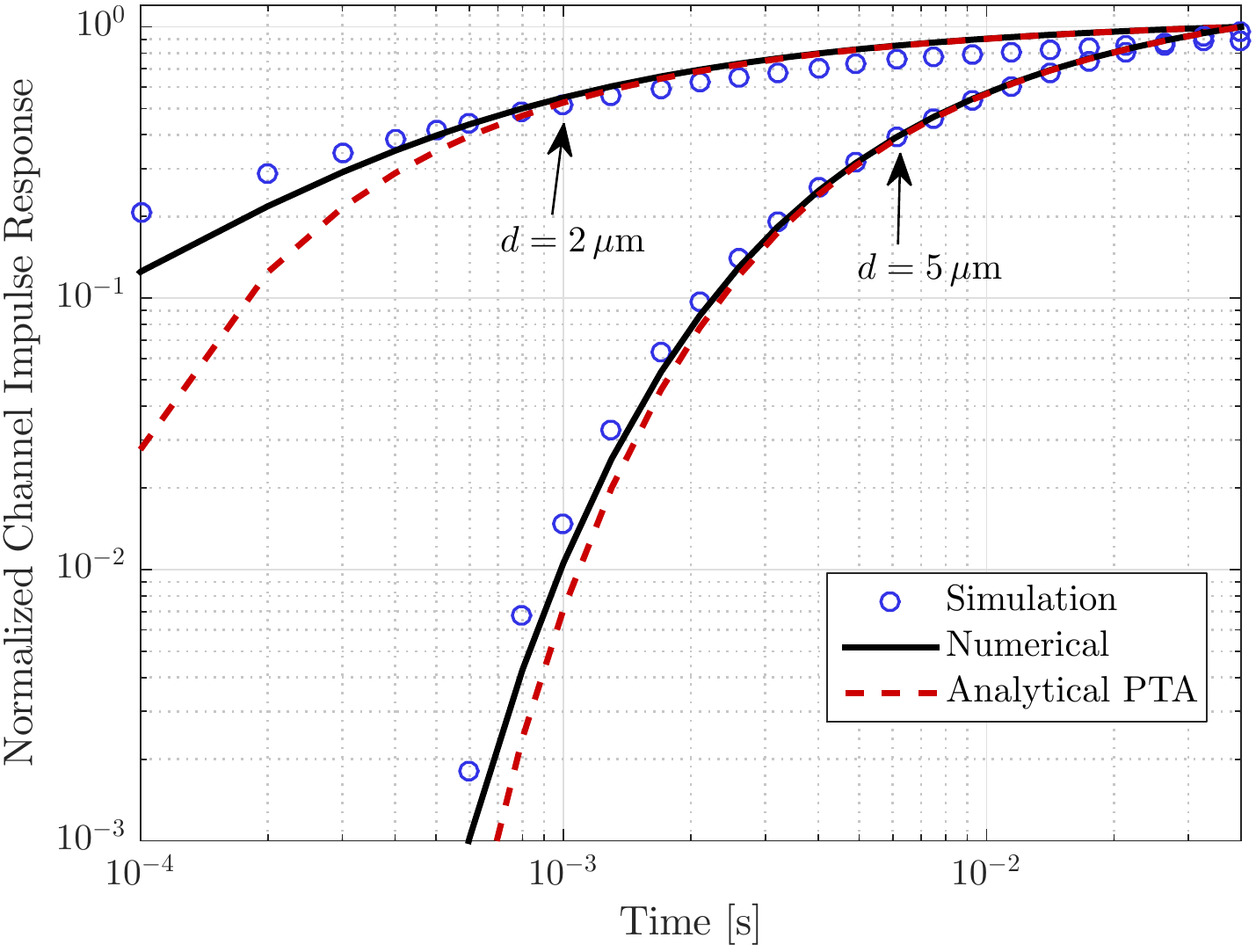}
	\caption{Absorbing channel impulse responses in the 3D environment as a function of time when the transmitter is a solid sphere. All curves are normalized to the maximum value of the corresponding analytical curve (solid black line).}
	\label{fig_3D_sphere}
\end{figure}

\section{Conclusions}
\label{sec_concl}

In this paper, we considered the channel impulse responses due to transmitters that release molecules into diffusing environments from a volume instead of a point. We derived the channel impulse responses for spherical passive and absorbing receivers in 1D and 3D environments due to a spherical transmitter. The impulse responses could be determined in closed form for the 1D environment. The accuracy of the derived impulse responses were supported by particle-based simulations. We referred to existing analysis, which assumes a point transmitter, as making the point transmitter assumption, and its accuracy was shown to generally improve with time. Future work will seek to obtain additional closed-form results to more accurately model the behavior of diffusing molecules that are released from finite volume transmitters.

\bibliography{2016_bsc_volume_tx}

\end{document}